\documentclass[prd, twocolumn, nofootinbib, floatfix]{revtex4-1}

\usepackage{amsmath}
\usepackage{graphicx}
\usepackage{dcolumn}
\usepackage{bm}
\usepackage{epsfig}
\usepackage{amssymb,latexsym,mathrsfs}
\usepackage{graphicx}
\usepackage{color}
\usepackage{hyperref}

\hypersetup{
    colorlinks=true,
    linkcolor=red,
    citecolor=blue,
} 

\newcommand{\be}{\begin{equation}}
\newcommand{\ee}{\end{equation}}
\newcommand{\bs}{\begin{split}} 
\newcommand{\bea}{\begin{eqnarray}}
\newcommand{\eea}{\end{eqnarray}}

\newcommand{\om}{\Omega_m}

\newcommand{\gmat}{G_{\rm matter}} 
\newcommand{\gm}{G_{\rm m}} 
\newcommand{\gl}{G_{\rm light}}

\newcommand{\sig}{\sigma} 
\newcommand{\fs}{f\sigma_8}
\newcommand{\fl}{f_\Lambda} 
\newcommand{\dg}{\delta G}
\newcommand{\dgm}{\delta G_{\rm m}}
\newcommand{\df}{\delta f}
\newcommand{\glam}{g_\Lambda}

\def\nn{\nonumber} 

\begin{document}

\title{Cosmic Growth Signatures of Modified Gravitational Strength} 
\author{Mikhail Denissenya${}^1$ and Eric V.\ Linder${}^{1,2}$} 
\affiliation{${}^1$Energetic Cosmos Laboratory, Nazarbayev University, 
Astana, Kazakhstan 010000\\ 
${}^2$Berkeley Center for Cosmological Physics \& Berkeley Lab, 
University of California, Berkeley, CA 94720, USA} 

\begin{abstract}
Cosmic growth of large scale structure probes the entire history of 
cosmic expansion and gravitational coupling. To get a clear picture 
of the effects of modification of gravity we consider a deviation in 
the coupling strength (effective Newton's constant) at different redshifts, 
with different durations and amplitudes. We derive, analytically and 
numerically, the impact on the growth rate and growth amplitude.  
Galaxy redshift surveys can measure a product of these through redshift 
space distortions and we connect 
the modified gravity to the observable in a way that may provide a 
useful parametrization of the ability of future surveys to test gravity. 
In particular, modifications during the matter dominated era can be treated 
by a single parameter, the ``area'' of the modification, to an accuracy 
of $\sim0.3\%$ in the observables. We project 
constraints on both early and late time gravity for the Dark Energy 
Spectroscopic Instrument and discuss what is needed for tightening 
tests of gravity to better than 5\% uncertainty. 
\end{abstract} 

\date{\today} 

\maketitle

\section{Introduction} 

Future galaxy redshift surveys will measure cosmic structure over an 
increasing volume of the universe, to higher redshift. One particular 
cosmological probe coming from the surveys is redshift space distortions, 
the angular dependence of galaxy clustering viewed in redshift space, 
a direct probe of the growth, and growth rate, of structure. This, in 
terms of the relation between the density and velocity fields, was 
identified as a test of cosmic gravity in an influential paper by Peebles 
\cite{0208037}. 

Measurements of redshift space distortion effects began to place 
significant constraints on the matter density 
\cite{peacock,hawkins,sdss04,sdss06,ross} and then were explicitly 
developed as tests of gravity \cite{07091113}. Galaxy redshift surveys 
observational constraints \cite{guzzo,08054789,gaztanaga} and further 
theoretical work treating the surveys \cite{wang,acquaviva,song,white} 
followed, as well as related techniques combining redshift space 
distortions with other probes, e.g.\ \cite{zhang}. This is now a 
common and significant part of modern survey cosmology. 

Modified gravity is a major possibility for the origin of current cosmic 
acceleration, and considerable effort is underway to understand how best 
to connect theoretical ideas with observational measurements in a clear 
and accurate way. Theories of modified gravity often have enough freedom 
that they can match a given cosmic background expansion, leaving a 
main avenue for distinguishing them from a cosmological constant or 
scalar field in terms of the alteration of the cosmic growth history. 
Within general relativity, the cosmic expansion determines the cosmic 
growth, but modified gravity allows deviations from this relation. 

On scales where density perturbations are linear, the scale dependence 
of modified gravity is generally negligible, with the time dependence 
the key factor. At the theory level, many time dependent functions can 
enter the action but phenomenologically for the growth of structure 
these can effectively be condensed to a modified Poisson equation relating 
the metric perturbations to the density perturbations, involving a 
single factor giving the gravitational coupling strength $\gmat(a)$, 
where $a$ is the cosmic expansion factor. 

Here our goal is to explore the connection between the deviations of 
$\gmat(a)$ from the general relativity case, where it is simply Newton's 
constant, and the observables from redshift surveys. One aim is to 
enable clearer understanding of the effects of modified gravity on 
growth measurements, without restriction to a particular theory. Another 
is to explore the possibility of a low order parametrization that could 
fruitfully be used to fit observational data to signatures of modification 
of gravity. 

In Sec.~\ref{sec:method} we review the modified Poisson equation and 
its influence on the evolution equation of density perturbations. After 
predicting analytically the effects on the cosmic growth rate and 
amplitude in certain limits, we explore the parameter space numerically 
in Sec.~\ref{sec:results}. We consider late time modifications in 
Sec.~\ref{sec:late}, demonstrating distinction between some broad 
classes, and derive projected constraints for future redshift surveys 
as probes of gravity in Sec.~\ref{sec:fisher}. We conclude in 
Sec.~\ref{sec:concl}.

\section{Cosmic Structure Growth and Gravity} \label{sec:method} 

Cosmic structure growth proceeds through a competition between 
gravitational instability -- an overdensity of matter attracting more 
matter under gravity -- and Hubble friction due to the cosmic expansion 
opposing growth. This gives the linear density perturbation evolution 
equation 
\be 
\ddot\delta+2H\dot\delta-\frac{3}{2}H^2\om(a)\gmat(a)\delta=0 \ , d
\label{eq:ddot} 
\ee 
where $\delta=\delta\rho_m/\rho_m$ is the matter overdensity, 
$H=\dot a/a$ is the Hubble parameter, where dot denotes a derivative 
with respect to cosmic time, $\om(a)=8\pi G_N\rho_m/(3H^2)$ is the 
dimensionless matter density as a fraction of the critical density, 
and $\gmat$ is the gravitational strength in units of Newton's constant 
$G_N$. 

The source term of the gravitational instability, the term proportional 
to $\delta$, arises from the modified Poisson equation relating the 
Newtonian gravitational potential $\psi$ to the density perturbation, 
\be 
\nabla^2\psi=4\pi G_N\gmat\rho_m\delta \ . 
\ee 
The growth equation as written assumes that the modification is purely 
to the gravitational coupling strength, that there are no nonminimal 
couplings of the matter sector to other sectors. For the rest of this 
article we abbreviate $\gmat$ as $\gm$. 

At high redshift, such as around last scattering of the cosmic microwave 
background (CMB), observations indicate that general relativity is an 
excellent 
description of gravity and so the initial conditions for the growth 
equation are taken to be unchanged. In the high redshift matter dominated 
universe, where $\om=1$ and $H^2=2/(3t)$, the solution for the growth 
is $\delta\propto a$. This makes it convenient to define a normalized 
growth factor $g=(\delta/a)/(\delta_i/a_i)$, where a subscript $i$ 
indicates an initial time in that epoch. 

The growth equation can then be written 
\bea 
g''&+&\left[5+\frac{1}{2}\frac{d\ln H^2}{d\ln a}\right]\frac{g'}{a}\nn\\  
&+&\left[3+\frac{1}{2}\frac{d\ln H^2}{d\ln a}-\frac{3}{2}\om(a)\gm(a)\right] 
\frac{g}{a^2}=0 \ , 
\eea 
where a prime denotes a derivative with respect to $a$. 
In order to focus on the impact of the modified $\gm$, we take the 
background expansion to be identical to that of $\Lambda$CDM, a flat 
matter plus cosmological constant universe. 

The mass fluctuation amplitude $\sigma_8$ is proportional to the growth 
factor $g$, but is difficult to extract from galaxy redshift surveys 
since the galaxy bias has a similar effect. The growth rate 
$f=1+d\ln g/d\ln a$ is of particular interest since it gives a more 
instantaneous sensitivity to the conditions at a particular redshift 
than the integrated growth that enters the growth factor. The observable 
from redshift space distortions (RSD), at the linear level, is the product 
$f\sigma_8$, or $fga\propto d\delta/d\ln a$. We will examine the impact 
of modified gravitational strength on $f$, $g$, and $\fs$. 

To build intuition for the physical interpretation of the later 
numerical results, let us begin with an analytic investigation. 
This can most fruitfully be done in terms of the growth rate equation, 
derived from the growth equation to be 
\be 
\frac{df}{d\ln a}+f^2+\left[2+\frac{1}{2}\frac{d\ln H^2}{d\ln a}\right]f 
-\frac{3}{2}\om(a)\gm(a)=0 \ . 
\ee 
Next consider the deviation in growth rate between the model with 
modified gravity and that without, i.e.\ standard $\Lambda$CDM: 
\bea
&&\frac{d(f-\fl)}{d\ln a}+\left[(f-1)^2-(\fl-1)^2\right]\nn\\ 
&+&\left[4+\frac{1}{2}\frac{d\ln H^2}{d\ln a}\right] (f-\fl) 
=\frac{3}{2}\om(a)\left[\gm(a)-1\right] 
\ . \label{eq:dffull} 
\eea 

Until dark energy begins to dominate, $f$ (and $\fl$) are close to one 
so we could neglect the square bracket involving the difference of the 
$(f-1)^2$ factors. Integrating the equation over $\ln a$, we find 
\bea 
\int_0^a &d&\ln a'\left\{\frac{d\df}{d\ln a'} 
+\left[4+\frac{1}{2}\frac{d\ln H^2}{d\ln a'}\right]\,\df\right\}\nn\\ 
&=&\frac{3}{2}\int_0^a d\ln a'\,\om(a')\left[G_m(a')-1\right] \ , 
\label{eq:finteq} 
\eea 
where $\df=f-f_\Lambda$ and $\dgm=G_m-1$. 
The first term on the left is a total derivative 
and $\df$ vanishes at early times so the contribution is simply $\df(a)$. 
Restricting to the matter dominated epoch, where $H^2\propto a^{-3}$ and 
$\om(a)=1$, yields 
\be 
\int_0^a d\ln a'\,\df\approx \frac{3}{5}\int_0^a d\ln a'\,\dgm 
-\frac{2}{5}\df(a) \ . \label{eq:farea} 
\ee 

This is a very interesting expression because recall the relation of 
growth rate to growth factor: 
\be 
g=a^{-1}\,e^{\int_0^a d\ln a'\,f(a')} \ ,
\ee 
and thus 
\be 
\frac{g}{\glam}=e^{\int_0^a d\ln a'\,[f(a')-\fl(a')]}= 
e^{\int_0^a d\ln a'\,\df(a')} \ . \label{eq:dgexpf} 
\ee 
If the deviations are small (recall that even for 10\% deviations in 
growth the difference between $e^x$ and the first order expansion $1+x$ 
is small, less than 0.5\%) then we can expand the exponential to 
get 
\be 
\frac{\delta g}{\glam}\approx \int_0^a d\ln a'\,\delta f(a') \ . 
\label{eq:dglinf} 
\ee 
That is, the growth factor deviation is approximately the area under 
the growth rate deviation curve, and Eq.~(\ref{eq:farea}) tells us 
this is closely related to the area under the gravitational strength 
curve. 

In particular, if the growth rate change from the gravitational 
modification has faded by the time at which the growth factor is 
evaluated, then we can neglect the $(2/5)\df(a)$ term in 
Eq.~(\ref{eq:farea}). Then the growth factor change is indeed 
proportional to the area under $\dgm$. We can be more precise by 
writing Eq.~(\ref{eq:dffull}) under the same assumptions of 
small deviations and matter domination and using the integrating 
factor method for solution. Then 
\be
\frac{d\df}{d\ln a}+\frac{5}{2}\df\approx \frac{3}{2}\dgm(a) 
\ee 
has the solution 
\be 
\df(a)\approx\frac{3}{2} a^{-5/2}\int_0^a d\ln a'\,a'^{5/2}\dgm(a') \ . 
\label{eq:dflone} 
\ee 
Substituting Eq.~(\ref{eq:dflone}) into Eq.~(\ref{eq:farea}) and 
Eq.~(\ref{eq:dgexpf}) gives 
\bea 
\frac{\delta g}{\glam}&\approx& \frac{3}{5}\int_0^a d\ln a'\,\dgm
-\frac{2}{5}\df(a)\\ 
&\approx& \frac{3}{5}\int_0^a d\ln a'\,\dgm 
\left[1-\left(\frac{a'}{a}\right)^{5/2}\right] \ . \label{eq:dgdf} 
\eea 
If the evaluation time $a$ is much after the epoch when $\dgm(a')$ 
is nonnegligible, then the square bracket quantity simply goes to one. 
In this situation the fractional growth factor deviation is just 
(three-fifths) the area under the gravitational modification curve. 

For the modified gravitational strength $\gm$ we want to use a 
parametrization that is tractable in terms of having only a few 
parameters, but that is consistent with the behavior of at least some 
theories of gravity. In particular, it should vanish at high redshift. 
To explore the signatures of modified gravity on growth, it is an 
advantage if $\dg$ is also fairly localized so we can explore the 
effect of deviations at different redshifts on growth during the 
observable epoch of $z\approx0-3$, where the redshift $z=a^{-1}-1$. 
That is, we want to build up our intuition and understanding of the 
connection between gravitational modifications and observables. 

We adopt the form 
\be 
\gm=1+\dg\,e^{-[(\ln a-\ln a_t)^2/(2\sig^2)]} \ , \label{eq:gmgaus} 
\ee 
where $\dg$ describes the amplitude of the deviation, $a_t$ the 
scale factor at which it peaks, and $\sig$ measures its duration. 
This fulfills the desired characteristics above, and is Gaussian 
in e-folds of expansion, $\ln a$. Such a peak gives similar results 
to the deviations seen in theories of modified gravity having multiple, 
competing terms in the Lagrangian, such as the Horndeski class; 
see Fig.~5 of \cite{160703113} for example. 

We emphasize that localization through use of a Gaussian is for 
clarity in interpretation; we derived above that the area under the 
gravitational modification curve was a key parameter, so one could 
equally well treat multiple Gaussians, or some other function, as 
long as it held to the assumptions used above. We also stress that 
the analytic arguments above were to guide intuition, and we do not 
assume matter domination at all redshifts, rather we take the expansion 
history to be that of $\Lambda$CDM. We discuss treatment of modifications 
at recent times in Sec.~\ref{sec:late} but again our main aim is to 
achieve some insight in understanding the signature of a deviation at a 
particular redshift on subsequent cosmic growth. 

Given a Gaussian, the area under the modification curve is easy to 
calculate and in particular if we are interested 
in the total growth factor to the present then we have 
\be 
\frac{\delta g_0}{g_{\Lambda,0}}\approx \frac{3}{5}\,{\rm Area}\approx 
\frac{3}{5}\sqrt{2\pi\sig^2}\dg 
\approx 1.5\,\sig\dg \ . \label{eq:dgapx} 
\ee 

To summarize, our analytic understanding is that the growth rate 
deviation $\delta f(a)$ should 
approximately trace $\dgm(a)$, with somewhat lower amplitude (e.g.\ at 
its peak, where $df/d\ln a=0$, $\df\approx(3/5)\dgm$), slightly shifted 
to later times due to the integral, and skewed to later times due to the 
$(a'/a)^{5/2}$ factor (or alternately due to that the magnitude of 
$df/d\ln a$ subtracts 
from the $\df$ term in the growth rate equation before the peak but 
adds to it afterward). The growth factor itself is in turn an 
integral over $f$, and if $\dgm$ and so $\df$ is sufficiently localized 
then at later times $\delta g$ should go to a constant offset 
proportional to the area under the gravitational modification curve, 
described by Eq.~(\ref{eq:dgapx}). 

In the next section we carry out a full numerical evolution of the 
cosmic growth and test our understanding of the signatures of this 
gravitational modification.

\section{Signatures in Growth Evolution} \label{sec:results} 

\subsection{Effects on observables} \label{sec:obs} 

\begin{figure}[htbp!]
\includegraphics[width=\columnwidth]{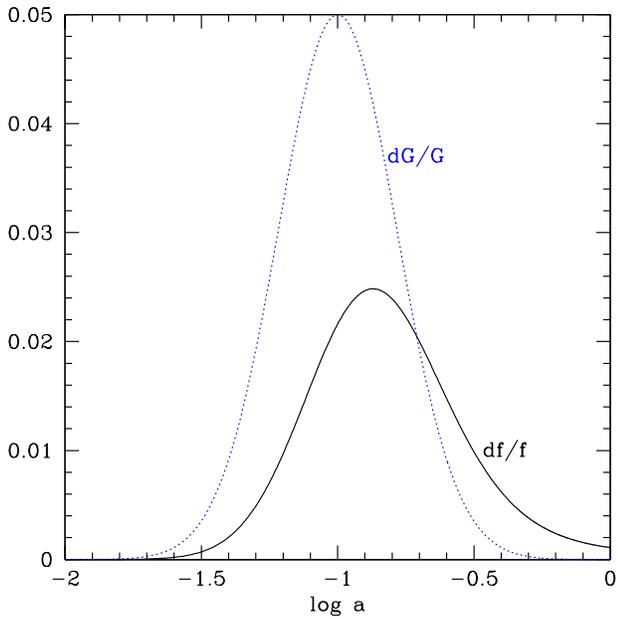} 
\caption{
A modification of the gravitational strength $\delta\gm$ propagates 
into an alteration of the cosmic growth rate $\df/f$. Here the modification 
has the parameters $\dg=0.05$, $a_t=0.1$, $\sig=0.5$; we plot it in 
$\log_{10} a$ rather than $\ln a$ for simplicity. The response of the 
growth rate is a slightly delayed, somewhat damped, near shadow of $\dgm$, 
due to 
the physics of the growth equation discussed in Sec.~\ref{sec:method}. 
} 
\label{fig:dfdglog10} 
\end{figure}

Taking a gravitational strength modification as a Gaussian in the 
expansion e-fold scale on top of the general relativity behavior, i.e.\ 
Eq.~(\ref{eq:gmgaus}), we solve numerically the growth evolution 
equation to obtain the cosmic growth rate $f$, growth factor $g$, and 
redshift space distortion amplitude $\fs$. Figure~\ref{fig:dfdglog10} shows 
the results for $\delta\gm=(\gm G_N-G_N)/G_N$ and $\df/\fl=(f-\fl)/\fl$ 
for the fiducial cosmology of a flat $\Lambda$CDM universe with present 
matter density $\om=0.3$.

Indeed our analytic expectations of the previous section are reasonably 
good. The quantity $\df/\fl$ is roughly Gaussian and slightly delayed 
from the gravitational strength perturbation. We can anticipate that 
if the redshift of the gravitational modification is moved closer to 
the present, or if its width is broadened, then the effect on $f$ might 
overlap the present. 

\begin{figure}[htbp!]
\includegraphics[width=\columnwidth]{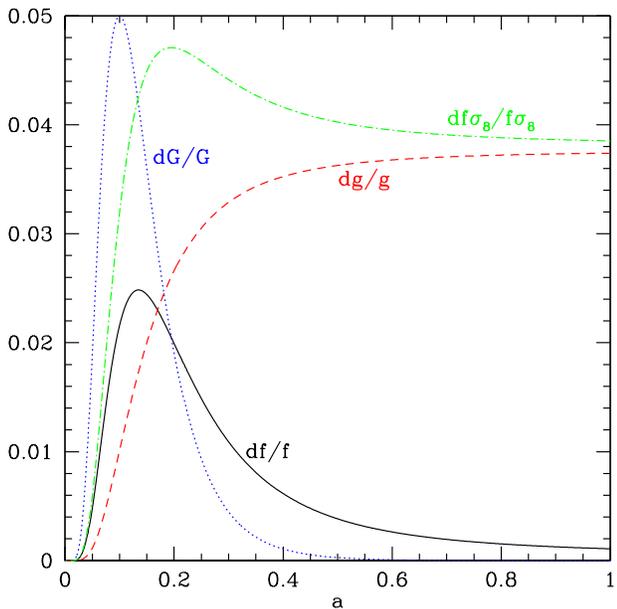} 
\caption{
The deviations of all the growth quantities in response to a 
gravitational modification are plotted vs $a$ for the same parameters 
as in Fig.~\ref{fig:dfdglog10}. In the recent universe the growth 
factor $g$ and RSD parameter $\fs$ go to constant offsets from the 
standard cosmology. 
} 
\label{fig:dfall} 
\end{figure}

Figure~\ref{fig:dfall} shows the responses of all the growth quantities, 
for the same parameters as Fig.~\ref{fig:dfdglog10} but plotted on a 
scale linear in expansion factor. Again we see that the analytic arguments 
hold fairly well: the growth factor approaches a constant offset at a 
time much later than the impulse of nonzero $\delta\gm$, basically when 
the delayed $\df/f$ also restores to the standard cosmology. Deviations 
in the RSD parameter $\fs$ acts like $\df/f$ at the beginning of the 
impulse, since $\delta\sigma_8\sim \delta g$ takes time to build (recall 
it is related to the area under the $\df$ curve), and like $\delta g$ 
at late times as $f$ restores to standard behavior. This shows that 
RSD are capable of testing gravity at all redshifts, from $z=0$ out to 
the epoch at which the modification peaks.

\begin{figure}[htbp!]
\includegraphics[width=\columnwidth]{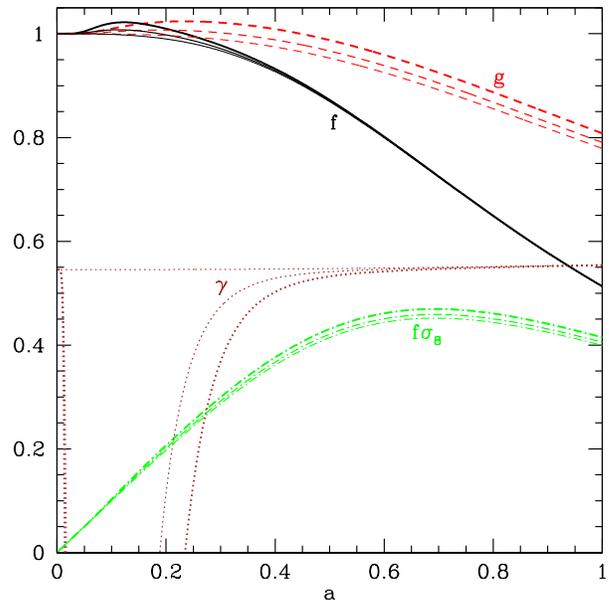} 
\caption{
The growth quantities are plotted vs $a$ for three cases: standard 
$\Lambda$CDM cosmology with general relativity (thin curves), with 
modified gravity with $\dg=0.02$ (medium curves), and with $\dg=0.05$ 
(thick curves). We also plot the gravitational growth index $\gamma$. 
} 
\label{fig:geff} 
\end{figure}

The growth quantities themselves, rather than the deviations from the 
general relativistic cosmology, are plotted in Fig.~\ref{fig:geff}. 
We see that the growth rate $f$ indeed restores to the standard 
evolution at recent times, and $g$ and $\fs$ suffer constant offsets. 
One subtlety is whether the growth amplitude is normalized at high 
redshift to the same initial conditions, i.e.\ cosmic microwave background 
power spectrum amplitude $A_s$, or at redshift 0, i.e.\ $\sigma_{8,0}$. 
We normalize to the CMB; if one instead normalized to $\sigma_{8,0}$ then 
the $\fs$ curves should be shifted vertically to agree at $a=1$.

Also plotted is the gravitational growth index $\gamma$, defined through 
$f(a)=\om(a)^\gamma$. This probes deviations from general relativity in 
the growth of matter perturbations \cite{lin05,lincahn} and we see its 
curves strongly pick up the gravitational modifications. The quantity 
$\gamma$ deviates from its general relativity $\Lambda$CDM value of 
$\gamma=0.55$ during the modification and then restores to it. Note that 
in the standard case $\gamma$ can be seen to be not perfectly constant 
at the value 0.55, but this is an excellent approximation to its behavior, 
especially in the integrated sense in which $\gamma$ enters the growth 
factor. During times of strengthened gravity, $\gamma$ gets smaller, 
i.e.\ the growth rate stays high even as the fractional matter density 
declines; when modified gravity increases the growth rate $f>1$, then 
$\gamma<0$. Thus $\gamma$ contains considerable information about 
modifications of gravity.

\subsection{Numerical vs analytic results} \label{sec:gaus} 

\begin{figure}[htbp!]
\includegraphics[width=\columnwidth]{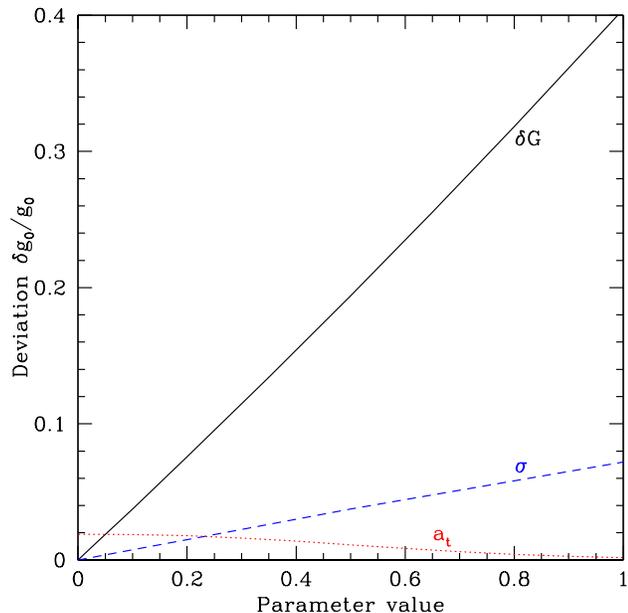} 
\caption{
The deviation in the growth factor to the present, $\delta g_0/g_0$, 
is plotted vs the value of each model parameter, varied one at a time 
around the fiducial $\{\dg,\sig,a_t\}=\{0.05,0.25,0.1\}$. 
} 
\label{fig:vary} 
\end{figure}

Now let us investigate how the variation of parameters within this 
model impacts the behavior of the cosmic growth variables, and 
conversely how sensitive the growth is at revealing characteristics of 
gravitational modification. Figure~\ref{fig:vary} varies the parameters 
of the model one at a time. We change the fiducial value of the width 
to $\sig=0.25$ since the narrower impulse gives clearer interpretation 
of the results.

The deviation in the growth factor increases nearly linearly with 
the amplitude $\dg$, and the same holds upon varying the duration $\sig$ 
of the modified gravitational strength. These are both just what is 
expected by Eq.~(\ref{eq:dgapx}), having their origin in the physics 
of the growth equation and within the approximation that the growth 
rate has substantially restored to the standard behavior by the present. 
The degree of linearity in the figure directly tests the validity of 
approximation. For the time of modification, $a_t$, we see that the 
growth is fairly insensitive to this provided it occurs early enough. 
As the modification peaks closer to the present, the inertia of the 
effect on the growth rate $f$ means that the influence on the growth 
factor $g_0$ diminishes toward zero. 

\begin{figure}[htbp!]
\includegraphics[width=\columnwidth]{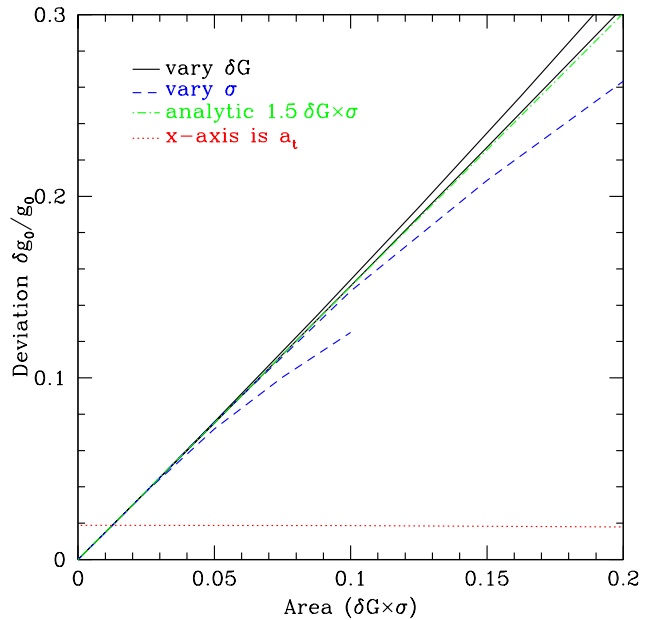} 
\caption{
The deviation in the growth factor to the present, $\delta g_0/g_0$,
is plotted vs the product $\dg\times\sig$, proportional to the area under 
the curve of gravitational modification. A linear fit is plotted as 
the green, dot-dashed line. The solid black curves show the results 
when varying the amplitude $\dg$, with $\sig=0.25$ and $\sig=0.05$ 
for the outer and inner curves respectively. The dashed blue curves 
show the results when varying the duration $\sig$, with $\dg=0.05$ 
and $\dg=0.1$ for the outer and inner curves respectively. The 
red dotted line is the same as in Fig.~\ref{fig:vary} and instead has 
the x-axis as the epoch of modification $a_t$, showing how little 
dependence there is on $a_t$ over the range $a_t=[0,0.2]$. 
} 
\label{fig:varyas} 
\end{figure}

Let us pursue this further, assessing the analytic approximation from 
Sec.~\ref{sec:method}. This predicts that the growth factor deviation 
should not only depend linearly on the amplitude and width of the 
modification, but that the key quantity is the area under the curve 
showing the departure of the gravitational strength from general relativity. 
We therefore plot in Fig.~\ref{fig:varyas} the growth factor deviation 
vs the product $\sig\,\dg$.

The linear fit to the growth behavior as a function of amplitude times 
width, or area, appears to be an excellent approximation, especially for 
the most 
observational viable values of the growth deviation (i.e.\ less than 
$\sim10\%$). The analytic prediction of Eq.~(\ref{eq:dgapx}) is so 
successful that it is interesting to understand what causes 
the slight deviations from linearity. 

As the amplitude $\dg$ varies to larger values of deviation, the 
behavior is to curve slightly up from linearity, more so for larger 
values of the width $\sig$. This arises from the increasing importance 
of the $(f-1)^2$ term in the growth rate Eq.~(\ref{eq:dffull}), or 
alternately the higher order expansion of the exponential in 
Eq.~(\ref{eq:dgexpf}), breaking linearity. Increasing $\sig$ further 
amplifies the effect of such high amplitudes on the growth factor. 

However, in the lower amplitude regime, increasing the duration $\sig$ 
causes the growth factor behavior to curve slightly down from linearity. 
This is due to the elongated persistence of the modification $\dg$ 
such that by the present where $g_0$ is evaluated the growth does not 
feel the full impact of the modification on $\df$, and hence $\delta g$. 
Essentially, one does not capture the entire area under the $\df$ 
curve in Eq.~(\ref{eq:dglinf}). Moving the modification epoch $a_t$ 
earlier would ameliorate this effect, while moving it later would 
exacerbate the nonlinearity. Recall though that late modifications 
give smaller deviations, all else equal, so the nonlinearity is less 
important in this case. 

To quantify the excellent agreement of the numerical results with the 
analytic prediction, to as high deviations in the growth as it does, 
note that the 
analytic relation works to 0.3\% in $g_0$ out 
to $\sig=1$, where the main effect is the modification persisting to 
the present. (Other parameters are held at their fiducial values.) 
At $\sig=0.5$ it is accurate at 0.01\%. The 
variation with respect to amplitude is much more forgiving, with 
accuracy of 0.02\% out to $\dg=1$. 

This close relation of the area of the modification 
to the growth factor deviation suggests that this 
quantity may play a useful role in parametrizing the gravitational 
modification and its signatures. We revisit this point later. 

\begin{figure}[htbp!]
\includegraphics[width=\columnwidth]{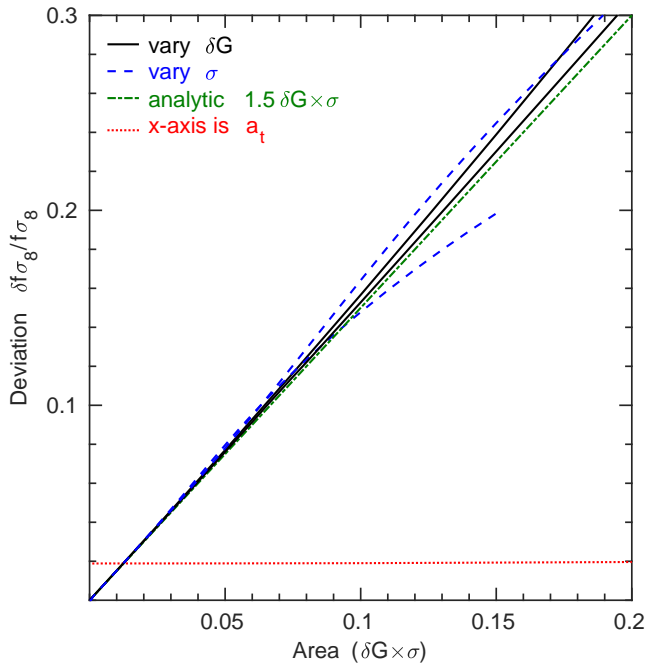} 
\caption{
As Fig.~\ref{fig:varyas}, but for $\delta\fs/\fs(z=0)$. Again, the 
analytic fit linear in area is an excellent approximation, especially 
for viable deviations less than $\sim10\%$. 
} 
\label{fig:dfsz0linear} 
\end{figure}

The same relation holds for $\delta\fs/\fs(z=0)$. One can readily see 
this analytically in that for small deviations this quantity is basically 
the sum of $\df/f$ and $\delta g/g$. Since $\df$ nearly vanishes by $z=0$, 
the behavior is nearly the same as for $\delta g/g$. In fact, the large 
$\sig$ deviations from linearity seen in Fig.~\ref{fig:varyas} for 
$\delta g/g$ are suppressed for $\delta\fs/\fs$ -- the analytic area 
relation works better because the suppression due to nonvanishing 
$\delta f(z=0)$ in Eq.~(\ref{eq:dgdf}), i.e.\ the correction factor in 
the square brackets, is counteracted by the $\delta f/f$ contribution 
to $\delta\fs/\fs$. This is evident in Fig.~\ref{fig:dfsz0linear}.

The RSD observable is accurately fit by the analytic area formula to 
0.4\% at $z=0$ and 0.6\% at z=1, out to $\sig=1$. However for 
$\sig=0.5$ the accuracy improves significantly to 0.1\% and 0.3\% 
respectively. For large amplitude modifications, the fit of $\fs$ 
weakens to the 3\% level for an extreme $\dg=1$ (which would 
entail a nearly 40\% deviation of $\fs$ from $\Lambda$CDM). 

\begin{figure}[htbp!]
\includegraphics[width=\columnwidth]{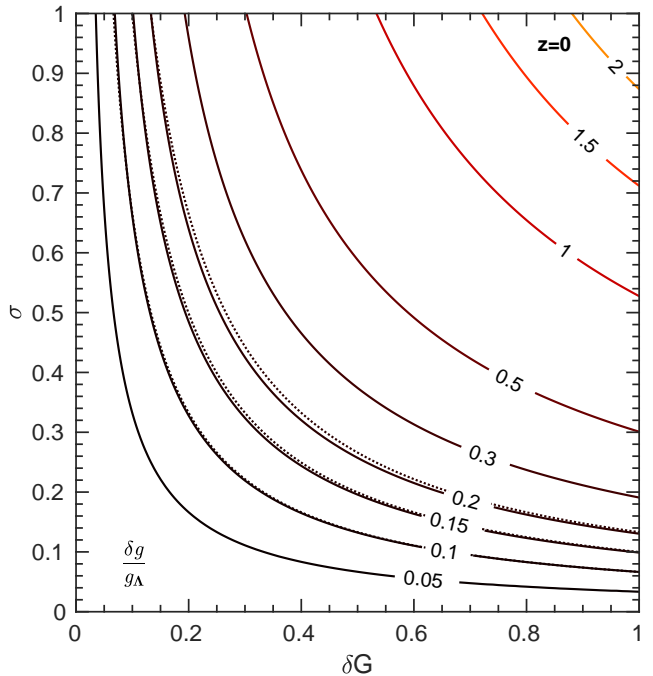} 
\caption{
Isocontours of $\delta g/g(z=0)$ are plotted in the $\sig$-$\dg$ plane, 
for fixed $a_t=0.1$. Dotted curves for the 0.05, 0.1, 0.15, and 0.2 level 
contours show the analytic, area prediction. 
} 
\label{fig:dgz0isosig} 
\end{figure}

\begin{figure}[htbp!]
\includegraphics[width=\columnwidth]{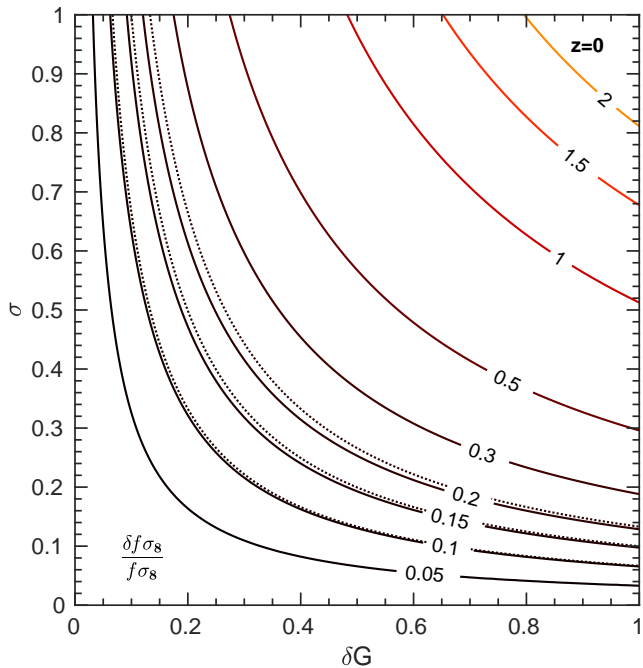} 
\caption{
As Fig.~\ref{fig:dgz0isosig} but for isocontours of $\delta\fs/\fs(z=0)$. 
} 
\label{fig:dfsz0isosig} 
\end{figure}

We can also illustrate the accuracy of the analytic approximation by 
showing the isocontours of the deviations in the growth factor and 
the redshift space distortion factor in Fig.~\ref{fig:dgz0isosig} 
and Fig.~\ref{fig:dfsz0isosig}, respectively. 
The shape of the $\fs$ contours is quite similar to those of the $g$ 
contours, and their level as well, as expected by the above reasoning. 
The dotted curves show the analytic, area 
approximation; this works superbly for the viable range of deviations 
less than about 10\%, and is quite reasonable even out to $\sim30\%$ 
deviations.

\begin{figure}[htbp!]
\includegraphics[width=\columnwidth]{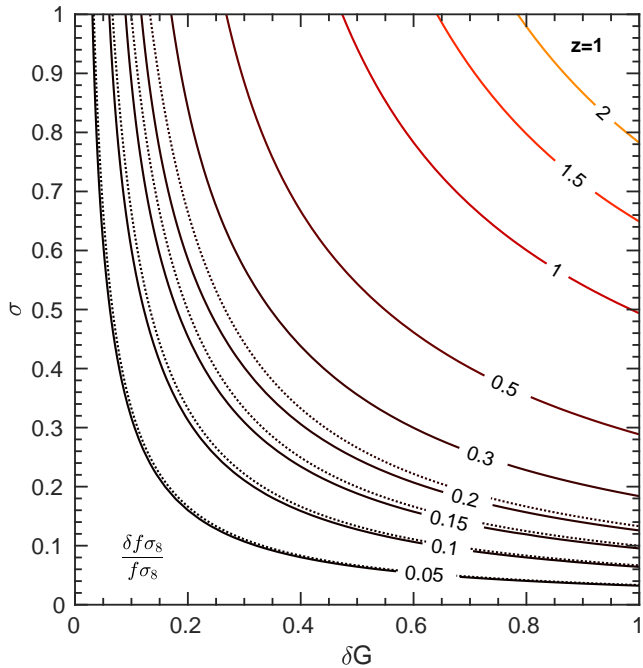} 
\caption{
As Fig.~\ref{fig:dgz0isosig} but for isocontours of $\delta\fs/\fs(z=1)$. 
} 
\label{fig:dfsz1isosig} 
\end{figure}

Since next generation galaxy redshift surveys aim to measure accurate 
redshift space distortions at $z\approx1$, we illustrate in 
Fig.~\ref{fig:dfsz1isosig} how well the area approximation holds for 
$\delta\fs/\fs(z=1)$. Since the $\delta f(a)$ term is less negligible 
at higher redshift, the integrand in Eq.~(\ref{eq:dgdf}) is suppressed 
somewhat from the pure area, but the analytic form is still quite accurate 
for viable deviations less than $\sim10\%$. 

Specifically, the analytic form for $g_0$ is good to 0.3\% everywhere 
along the 10\% deviation curve, and to 1\% for $g(z=1)$. However, if 
we restrict to $\sig\lesssim0.5$ then the accuracy improves to 0.1\% 
and 0.2\% respectively. For the RSD observable $\fs$ the accuracy is 
0.7\% at $z=0$ and 1\% at $z=1$, tightening to 0.4\% and 0.6\% for 
$\sig\lesssim0.5$. Recall that the extreme 1\% case corresponds to a 
change in $\fs$ of only 0.004, beyond even next generation survey 
precision. 

We find there is also little sensitivity to the value of the gravitational 
modification epoch $a_t$, as long as $a_t\lesssim0.25$ ($z_t\gtrsim 3$) 
and the observational 
quantity deviations are viably small (less than $\sim10\%$).

\subsection{Extended modifications} \label{sec:extend} 

\begin{figure}[htbp!]
\includegraphics[width=\columnwidth]{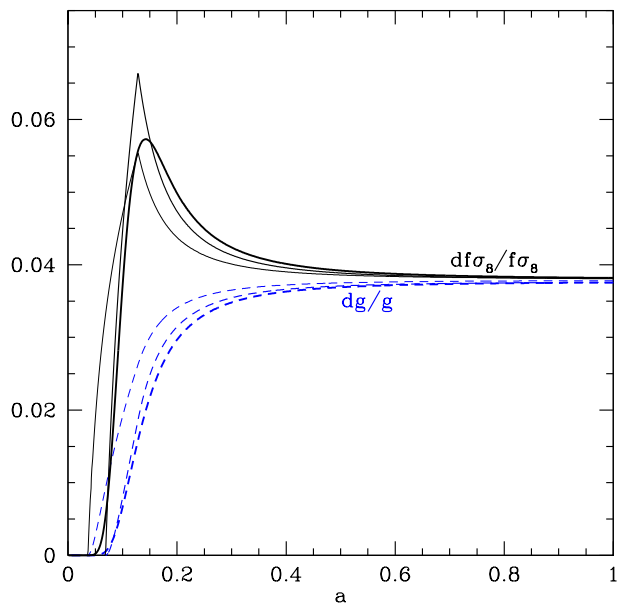}
\caption{
The deviation in the growth factor evolution, $\delta g/g$ (dashed blue 
curves), and the RSD observable evolution, $\delta\fs/\fs$ (solid black 
curves), are plotted for three 
different models. The Gaussian model, with $\dg=0.1$, $\sig=0.25$, $a_t=0.1$ 
(thick curve), a box model with $\dg=0.1$ and e-fold duration 
$\Delta\ln a=0.63$ ending at $\ln a_{GR}=\ln a_t+\sig$ (medium curve), 
and a box model with half the amplitude and twice the duration, ending at 
the same scale factor (thin curve), all have the same area under the 
gravitational modification, and hence nearly the same growth evolution 
for $a\gtrsim0.3$. 
}
\label{fig:box} 
\end{figure}

While Sec.~\ref{sec:method} showed analytically that the late time growth 
evolution should depend on the area of the modification, and we demonstrated 
this numerically in Sec.~\ref{sec:gaus} for a localized modification of 
various amplitudes and widths, we now illustrate this for extended early 
modifications. Figure~\ref{fig:box} plots the deviations in the growth 
factor $g(a)$ and the RSD observable $\fs(a)$ due to 
three different forms for the gravitational modification: a Gaussian as 
used in earlier plots, a box function with the same peak amplitude but 
the width adjusted to match the same area, and a box function with half 
the amplitude but twice the width (in e-fold, i.e.\ $\ln a$, units), 
so it also has the same area.

We see that indeed the quantities $\delta g/g$ and $\delta\fs/\fs$ are 
each nearly identical between models for $a\gtrsim0.3$ despite the 
gravitational modification 
redshift dependences being very different, just their area being preserved. 
Quantitatively, the deviations between the growth factors $g(a)$ for 
the Gaussian modification and the box modification are less than 0.1\% 
(0.25\%) for $a\ge0.25$ for the box with the same peak amplitude 
(half the peak amplitude, twice the duration). For the RSD observable 
$\fs(a)$, the corresponding deviations are 0.15\%, 0.35\%. 
This lends credence to the concept that an acceptable parametrization of 
matter dominated 
era gravitational modifications (such as are predicted by many theories 
involving multiple terms in the Lagrangian, e.g.\ in the Horndeski 
class of gravity) is a single parameter corresponding to the area of 
the modification, for matter growth observables. 

In the next section we explore late time modifications, where no such 
simplification is evident.

\section{Late time modifications} \label{sec:late} 

From the growth evolution equations, we see there is no physical 
expectation that the area property should hold for ``late time'' 
gravitational modifications once matter domination wanes. Therefore 
a simple parametrization of such gravitational modifications, and 
their effect on cosmic growth, is not obvious. To explore the 
diversity of behaviors, we take a phenomenological ansatz describing 
three basic modifications during the recent universe: one constant 
with scale factor, one increasing, and one decreasing. 

Specifically, we investigate 
\bea 
\dg(a)&=&\dg_c\\ 
\dg(a)&=&\dg_r\,a^s \label{eq:dgarise}\\ 
\dg(a)&=&\dg_f\,a^{-s} \ , \label{eq:dgafall}  
\eea 
over the range $a=[0.25,1]$, and zero otherwise (since we have 
treated the matter domination era gravitational modifications 
separately). We can choose the amplitudes of the constant, rising, 
and falling modifications to match in area, e.g.\ 
$\int_{0.25}^1 d\ln a\,\dg(a)=0.05\ln 4$ corresponding to the constant 
case with amplitude $\dg_c=0.05$. Then 
\bea 
\dg_r&=&0.05\ln 4\ \frac{s}{1-4^{-s}} \label{eq:dgrise}\\ 
\dg_f&=&0.05\ln 4\ \frac{s}{4^s-1} \ . \label{eq:dgfall} 
\eea 
We consider $s=3$. 

\begin{figure}[htbp!]
\includegraphics[width=\columnwidth]{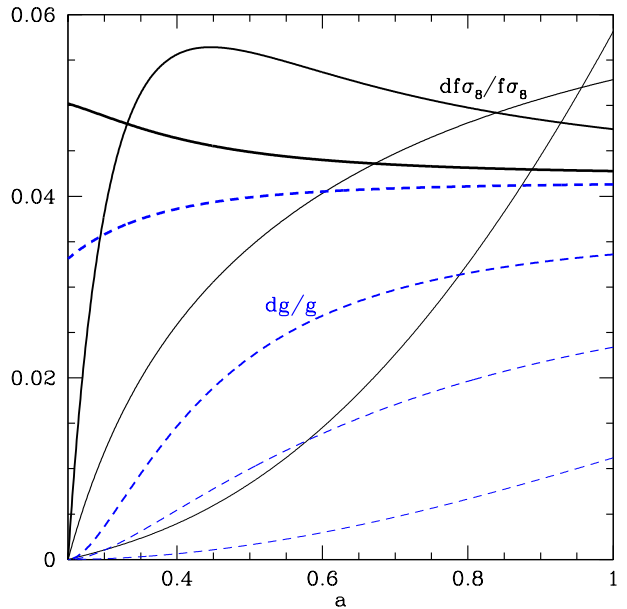}
\caption{
The deviation in the growth factor evolution, $\delta g/g$ (dashed blue 
curves), and the RSD observable evolution, $\delta\fs/\fs$ (solid black 
curves), are plotted for three 
late time modification models, as well as one early time modification 
model. All models have the same area under the gravitational modification. 
The thin/medium/thick curves correspond to the late time 
rising/constant/falling models with power law index $s=3$, 0, $-3$ 
between $a=0.25$--$1$. The thickest curve is the standard early time 
Gaussian modification with $\delta G=0.05$ and $\sig=0.553$ to match 
the area. Note that unlike in Fig.~\ref{fig:box}, curves with identical 
late time modification areas can be readily distinguished. 
}
\label{fig:late} 
\end{figure}

Figure~\ref{fig:late} exhibits the impact on the growth factor and 
growth rate evolution. Despite the gravitational modification areas 
being identical, the behaviors of the observables are quite different, 
losing the immunity to variation of the modification parameters (under 
conserved area) found for the early time modifications, e.g.\ in 
Fig.~\ref{fig:box}. The late time modifications also give signatures 
in the growth observables distinct from that of early time modifications. 
The thickest curve in Fig.~\ref{fig:late} is for the usual early Gaussian 
form with $\dg=0.05$, $a_t=0.1$, and $\sig=0.553$, with the value of 
$\sig$ chosen to match the area constraint.

The differences in the shapes, i.e.\ the evolution, of the observables 
indicate that galaxy survey measurements have the potential to distinguish 
between these classes of rising/constant/falling modifications, and 
moreover between late and early modifications, if the measurements extend 
beyond $z\approx1.5$. The greatest similarity between early and late time 
variations is for the falling class, as expected since this gives the 
greatest modification at smaller $a$ like the early time class. Even so, 
by $a\lesssim0.4$ the behavior of $\fs(a)$ is significantly different 
between the two. In the growth factor this is even clearer, for 
$a\lesssim0.5$, but the possibility of evolving galaxy bias makes this 
less dependable. In the next section we quantify the ability to probe 
gravity with future galaxy redshift surveys measuring $\fs(a)$ through 
redshift space distortions. 

First, though, let us combine the results of the two sections on 
early and late gravitational modifications. A general modification 
could be viewed as the sum of these two, giving a more arbitrary 
$\gm(a)$. This could well be nonmonotonic, as is common in theories 
of gravity with multiple terms, such as the Horndeski class, and 
seen in Fig.~5 of \cite{160703113} for example. We are particularly 
interested in how the sum of early and late modifications translates 
into the impact on cosmic growth quantities such as the growth factor 
and RSD observable. 

\begin{figure}[htbp!]
\includegraphics[width=\columnwidth]{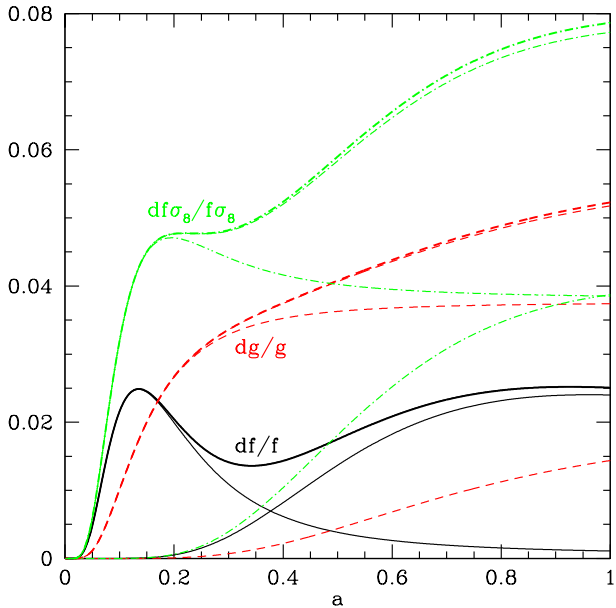} 
\caption{
The deviations of all the growth quantities in response to a nonmonotonic 
gravitational modification consisting of two overlapping Gaussians are 
plotted vs $a$. The thick curves show the results, while the thin curves 
show the behaviors for each individual Gaussian modification, and their 
sum. The full results are very close to those for the sum of the individual 
contributions to $\gm(a)$. 
} 
\label{fig:gaus2} 
\end{figure}

We choose a nonmonotonic deviation $\dgm(a)$ given by the sum of two 
Gaussians, 
one early ($a_t=0.1$), one late ($a_t=0.67$), with widths $\sig=0.5$ 
such that they overlap, very similar to what is seen in 
Fig.~5 of \cite{160703113}. Figure~\ref{fig:gaus2} shows the results 
for $\delta g/g$, $\delta f/f$, and $\delta\fs/\fs$. The overall effect 
is very close to the sum of the effects from the individual contributions 
to the modification.

We can understand this analytically. Consider Eq.~(\ref{eq:finteq}), 
with or without the integrals. 
The right hand side can be written as the sum of individual contributions 
to $\gm(a)$, and if we write $\delta f$ on the left hand side as the 
sum of the corresponding $\delta f_i$, then we see that the equation holds 
both for each contribution and for their sum. The reason this works is 
that we linearized the full Equation~(\ref{eq:dffull}) for the growth rate. 
For small (i.e.\ observationally viable) deviations of $f$ from $\fl$ 
-- one does not require $f$ close to 1 as in the matter dominated regime, 
just that $f$ is close to $\fl$ -- then the term $[(f-1)^2-(\fl-1)^2]$ 
has negligible effect. 

This sum rule propagates to the growth factor through 
Eq.~(\ref{eq:dgexpf}), and again to linear order in the growth deviations 
the deviation in the sum of modification contributions is simply the 
sum of the deviations, $\delta g_{1+2}=\delta g_1+\delta g_2$. For 
$\fs$ the situation is slightly more involved as the product of 
$\delta f$ and $\delta g$ enters, but the summation still works well. 
These are all evident from Fig.~\ref{fig:gaus2} where the summed 
$\delta f$ curve cannot be distinguished from the full result, the 
$\delta g$ curves are barely distinguishable and the $\fs$ curves are 
close. The maximum deviations are 0.00003, 0.0005, and 
0.001 for the three growth quantities, well below observational uncertainty. 

The property that gravitational modifications can be treated as 
sums of their contributions, say early and late modifications, with 
respect to the growth observables, has a significant and useful 
implication. It indicates that we can partially solve the parametrization 
problem: for a nearly arbitrary gravitational modification history we can 
accurately treat the gravitational modifications during 
the matter dominated epoch as in Sec.~\ref{sec:results}, with a single 
parameter corresponding to area, and we are then left with how to 
parametrize the late time modifications, from $z=0-3$, say. One 
possibility is to try to at least distinguish rising/constant/falling 
behaviors over this more restricted range, which adds two more parameters. 
We explore the possibility of such identification through their distinct 
signatures on observational quantities in the next section.

\section{Future constraints on gravity} \label{sec:fisher} 

To estimate constraints on the gravitational modifications from 
measurements of the redshift space distortion parameter $\fs(a)$ 
from future galaxy redshift surveys, we employ the Fisher information 
matrix formalism. This looks at the sensitivity of the observable to 
each model parameter and takes into account similarities in the 
response, i.e.\ covariances, to translate a given precision in data 
to a parameter constraint. 

For the future RSD measurements, we adopt the precisions given for 
the Dark Energy Spectroscopic Instrument (DESI) in Tables~2.3 and 
2.5 of \cite{161100036} for $\fs(a)$ between $z=0.05$--1.85. We use 
only linear scales, out to $k_{\rm max}=0.1\,h$/Mpc, since the impact 
of gravitational modification, and in particular its scale dependence, 
on $\fs$ beyond this is not clearly known. We will show in one case 
that assuming linear theory results hold out to $k_{\rm max}=0.2\,h$/Mpc 
does not yield significant improvement because of covariances; a robust 
treatment of scale dependence may well break this impasse. Also, we 
do not include the growth factor $g(a)$ since it is degenerate with 
galaxy bias in the linear regime. Again, improvements may be made with a 
robust treatment of bias at higher wavenumbers. 

Within a flat $\Lambda$CDM background, the parameters affecting $\fs$ 
are the matter density $\om$, the mass fluctuation amplitude $\sig_8$, 
and the gravitational modifications in the matter dominated era, 
$\dg_{\rm hi}$, and more recently, $\dg_{\rm lo}$. We have shown, 
both analytically 
and numerically, that gravitational modifications during the matter 
dominated era have an influence on $\fs$ at later times well approximated 
by a fractional offset, i.e.\ a multiplicative factor, proportional to 
the area under the modification curve. Thus, the area (or 
$\dg_{\rm hi}\times\sig_{\rm hi}$) is degenerate with $\sig_{8,0}$, 
the present 
value of the mass fluctuation amplitude. That is, one could obtain the 
same amount of structure with an intrinsically low amplitude and extra 
growth or a higher amplitude and less growth. Therefore we combine these 
into a parameter $S=[1+(3/5){\rm Area_{hi}}]\,\sig_{8,0}$. 

For lower redshift gravitational modifications, we investigate the 
three classes discussed in Sec.~\ref{sec:late}. These have two parameters, 
an amplitude and power law index. The fiducial values for the calculation 
are $\om=0.3$, $S=0.85$ (e.g.\ $\sig_{8,0}=0.82$, $\dg_{\rm hi}=0.05$, 
$\sig_{\rm hi}=0.5$), with the low redshift gravitational modification 
corresponding 
to $\dg_{\rm lo}=0.05$ when $s=0$, i.e.\ constant, and otherwise given 
by Eqs.~(\ref{eq:dgarise})--(\ref{eq:dgfall}) for $s=3$. 

\begin{figure}[htbp!]
\includegraphics[width=\columnwidth]{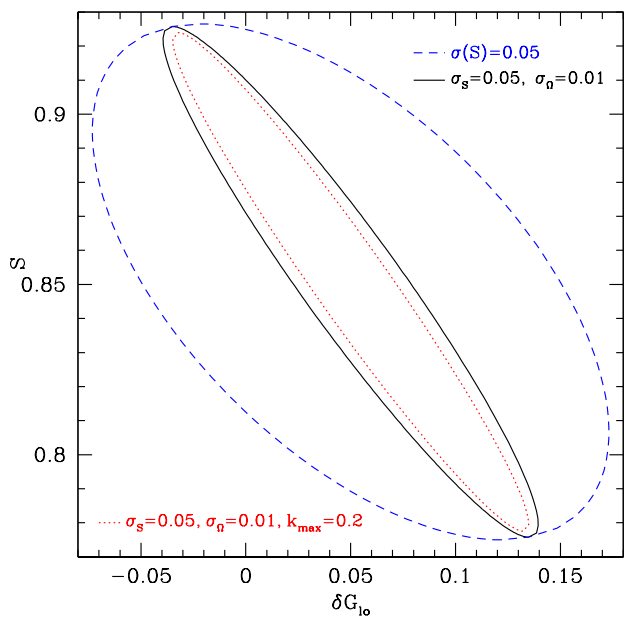} 
\caption{
Future constraints on the low redshift gravitational modification 
$\dg_{\rm lo}$ and the mass fluctuation amplitude $S$ combining 
$\sig_{8,0}$ and the high redshift modification are plotted as 
joint 68\% confidence contours, marginalizing over the other parameters. 
This uses the falling class $\dg\propto a^{-3}$ and we can see the 
strong covariance with $S$. The dashed blue contour applies an external 
prior $\sig(S)=0.05$, the solid black contour adds a prior $\sig(\om)=0.01$, 
and the dotted red contour sharpens the measurement precision $\sig(\fs)$ 
by roughly a factor two by extending to $k_{\rm max}=0.2\,h$/Mpc (see 
\cite{161100036}). 
} 
\label{fig:losfall} 
\end{figure}

Figure~\ref{fig:losfall} shows the results for the falling case, where 
the modification is decreasing from higher redshifts. From 
Fig.~\ref{fig:late} we see that this model has the highest amplitude 
effect on $\fs$, but also a fairly constant amplitude for $z\lesssim2$, 
which could lead to covariances. Indeed, that is what the results show. 
An overall diagonally oriented covariance between $dg_{\rm lo}$ and $S$ 
is seen, 
with the thickness of the confidence contour sensitive to the uncertainty 
in $\om$. Recall that the source term in the growth equation~(\ref{eq:ddot}) 
involves the product $\om G\delta\sim\om G\sigma_{8,0}$.

We can view the priors 
on $\om$ and $S$ as coming from information in the galaxy survey besides 
the growth rate measurements, or from other experiments. Due to the 
diagonally oriented covariance, the rule of thumb is that the uncertainty 
on the low redshift gravitational modification will be of order the 
uncertainty on $S$, i.e.\ the overall mass fluctuation amplitude, 
$\sig(\dg_{\rm lo})\sim\sig(S)$. 

Improving the precision of the $\fs$ measurements from this level, e.g.\ 
by going to higher $k_{\rm max}$ -- assuming no new scale dependence to 
bring degeneracies, does not significantly tighten the constraints due 
to this covariance. The degeneracy needs to be broken, by direct 
measurement of the mass fluctuation amplitude (e.g.\ by the CMB at high 
redshift and by galaxy clusters or weak gravitational lensing at lower 
redshift, or by further information within the galaxy survey itself). 

\begin{figure}[htbp!]
\includegraphics[width=\columnwidth]{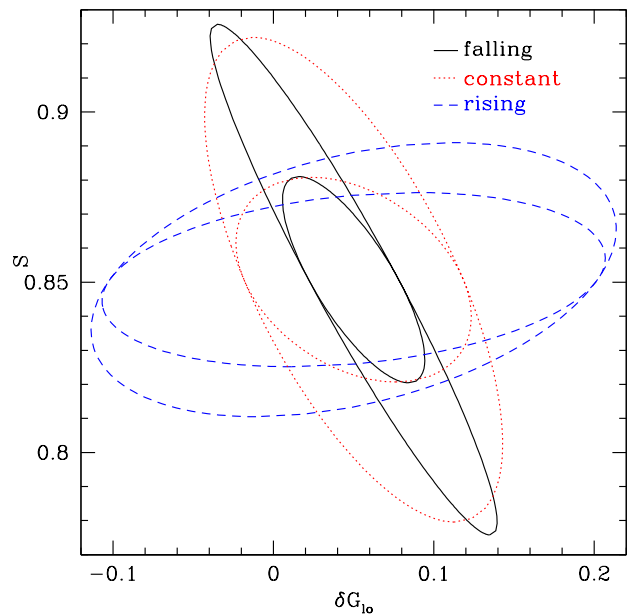} 
\caption{
As Fig.~\ref{fig:losfall} but contrasting the constraint for the three 
classes of falling, constant, and rising modified gravitational strength. 
The outer solid black contour corresponds to that from 
Fig.~\ref{fig:losfall} for the falling class. Inner contours for each 
class use a tightened external prior of $\sig(S)=0.02$ to show the impact 
the covariance of this parameter with $\dg_{\rm lo}$ has on the gravity 
constraint. 
} 
\label{fig:los05om01all} 
\end{figure}

We compare the different classes of low redshift gravitational 
modification in Fig.~\ref{fig:los05om01all}. The results show an 
interesting interplay between the shapes and amplitudes of the $\fs$ 
deviations exhibited in Fig.~\ref{fig:late}. Recall all three cases 
have same integrated gravitational modification, i.e.\ ``area''.

As stated, the falling case has the greatest amplitude of $\fs$ deviation 
during the bulk of the redshift range observed, but with a shape not 
well distinguished from a constant offset such as $S$ gives. Therefore 
it has a diagonally oriented covariance. The rising case has a shape 
distinct from 
the standard cosmological parameters, and thus has relatively little 
covariance with them, but also a low amplitude of deviation during the 
important redshift range, giving weaker constraints on $\dg_{\rm lo}$: 
i.e.\ a more horizontal, and broader confidence contour. The constant 
case is somewhat in between, with less covariance but also a low 
amplitude. We also investigate the case of the Gaussian low redshift 
modification of Fig.~\ref{fig:gaus2} and as expected it also has little 
covariance with $S$ (not shown), but is over a restricted redshift range; 
its constraints fall between the 
constant and rising cases. The 5-10\% constraints on gravitational 
modifications we find are comparable to those of \cite{12120009}, 
which used $\gm$ piecewise constant in redshift bins. 

We can understand the 5-10\% limit, at least within a factor of a few, 
by considering the expressions from Sec.~\ref{sec:method} relating $\dg$ 
and the measurement precision $\delta\fs/\fs$. The quantity 
$\delta\fs/\fs$ involves $\delta g/g$ and $\delta f/f$, both of which 
are integrals over $\dg(a)$. Basically there is a linear functional 
relation between $\dg$ and $\delta\fs/\fs$, so for unmarginalized 
uncertainties $\sig(\dg)\sim\sig(\delta\fs/\fs)$. An experiment with 
a precision of 2\% in $\delta\fs/\fs$ (as DESI achieves over a certain 
redshift range) should deliver an unmarginalized constraint on $\dg$ 
of the same order. 

In a bit more detail, the integral over $\dg(a)$ 
[cf.~Eq.~(\ref{eq:dgdf})] outside the matter dominated era is weighted 
by the matter density $\om(a)$ and a dilution factor 
of $(a^4 H)^{-1}$ (which gives the $a^{-5/2}$ in the matter dominated era). 
Furthermore, there are multiple measurements of $\fs$ at various redshifts, 
which reduces the uncertainty on $\dg$. We can incorporate all these 
effects into the following illustrative approximation, 
\be 
\sig(\dg)\approx\sig\left(\frac{\delta\fs}{\fs}\right)\,\frac{1}{N_{\rm eff}} 
\ , 
\ee 
where $N_{\rm eff}$ is the effective, weighted number of e-folds going 
into the integral over $\dg(a)$. The time when $\dg(a)$ is significant 
gives a tradeoff between the weighting factors and the persistence 
(that an early deviation has a lasting effect in $\fs$) such that 
$N_{\rm eff}$ is largest for the falling case and smallest for the 
rising case. This also interplays with the redshift dependence of the 
measurement precision $\sig(\delta\fs/\fs)$, though DESI has 2-3\% 
precision over a substantial redshift range. The unmarginalized 
uncertainties $\sig(\dg)=0.0071$, 0.011, 0.036 for the falling, constant, 
rising cases, respectively, not substantially different from the 2\% 
measurement 
precision. For the marginalized uncertainty, one must fold in the 
covariances, especially with $S$; since the falling case has more 
degeneracy and the rising case has little, the final results all end 
up in the 5-10\% constraint range. 

In all cases the power law of the modification scale factor dependence 
is poorly determined, of order $\sig(s)\approx4$. This means that one 
cannot distinguish between the classes. Late time modification of 
gravity thus remains a challenging subject, both theoretically and 
observationally. Addition of CMB and gravitational lensing data will 
help, because they both also depend on $\gmat$; however, they depend on 
the gravitational coupling for light deflection, $\gl$, as well so an 
analogous formalism or parametrization scheme is 
required for this quantity. One particularly interesting future 
prospect is the kinetic Sunyaev-Zel'dovich effect used to measure the 
velocity field, probing $H\fs$ (see, e.g., \cite{160401382}). 
Understanding of galaxy bias and scale 
dependent growth will also be useful if we aim to go beyond 5\% tests 
of gravity.

\section{Discussion and Conclusions} \label{sec:concl} 

A new generation of galaxy redshift surveys will vastly increase the 
volume and depth of the universe over which we measure the cosmic 
growth history. This brings with it the ability to test the foundations 
of gravity and look for modifications of general relativity, i.e.\ to 
confront possible extensions with observations. This can be done model 
by model, or one can seek general signatures in the observations that 
point to properties of the unknown theory of gravitation. We took the 
latter approach, 
considering the effective Newton's constant -- the gravitational strength 
entering into the Poisson equation for the growth of structure, called 
$\gmat$. 

This governs the growth rate, amplitude, and the product of these that 
enters the redshift space distortion observable. Remarkably, we found 
that a modification taking place at any time during the matter dominated 
era, i.e.\ $z\gtrsim3$, could be parametrized in terms of a single number 
-- the area under 
the deviation curve $\dgm(a)$ with respect to e-fold $\ln a$. We derived 
this analytically and demonstrated it numerically. Whether the deviation 
is localized, extended, or nonmonotonic, the area approximation reproduces 
the growth observables to $\lesssim0.3\%$ in the growth factor and 
$\lesssim0.6\%$ in the RSD quantity $\fs(z\approx1)$ in most cases, 
better than the measurement precision of next generation surveys. 

Such an accurate, derived parametrization dramatically simplifies the 
task of comparing gravitational modifications to cosmic growth 
observations. Recall that many gravity theories, in particular much of 
the Horndeski class, predict such matter era variations. Furthermore, 
we demonstrated that the full gravitational modification history could 
be accurately treated by the sum of the separate  matter dominated era 
(``early'') and late impacts on the growth quantities. For example, 
this sum reproduces the exact RSD observable $\fs$ to within 0.001, 
well below the statistical uncertainty. Combined with the previous 
result, this reduces the treatment of the entire modified gravity 
history to one number (the area from the early modification) plus 
a description at $z\lesssim3$. 

For the late time description we considered three classes, where the 
gravitational modification was rising, constant, or falling with 
scale factor over the range $z=0$--3. We showed that these had distinct 
effects on the growth observables. However, covariances with other 
cosmological parameters needed to be taken into account, so we performed 
a Fisher information analysis using the measurement precisions on 
$\fs(a)$ baselined for the upcoming DESI galaxy redshift survey over 
the range $z=0$--1.9. 

The projected constraint analysis showed that DESI could achieve 
gravitational modification amplitude estimation at the 5-10\% level, 
with the limiting factor being the covariances, particularly with the 
mass fluctuation amplitude $\sig_{8,0}$ and also the matter density 
$\om$. Also, the rising/constant/falling classes could not be reliably 
distinguished. 

In order to obtain a significant improvement, future galaxy surveys 
would need to strengthen its measurement precision on $\fs(a)$ 
to below 2\%, or additional probes of gravity (such as lensing, CMB, 
and galaxy clusters) or 
tighter external priors on covariant parameters need to be implemented. 
Even 1\% measurements of $\fs$ across the entire $z=0$--1.9 range give 
2.6\%, 3.0\%, 4.4\% constraints of gravity for the three classes. 

Testing gravity experimentally, and connecting the observations to 
theory, is a challenging subject. In one sense, this work has ``solved'' 
the problem for $z=3$--1000 and only left the last 1.5 e-folds of 
cosmic history lacking a clear connection. That is less than 
satisfactorily enlightening, however, and the remaining work on how 
to effectively and practically parametrize the late time gravitational 
modifications is substantial. Other aspects of gravity, such as how to 
characterize $G_{\rm light}$ for light propagation and other modifications 
affecting gravitational wave propagation, also require future work.

\acknowledgments 

This work is supported in part by the Energetic Cosmos Laboratory and by 
the U.S.\ Department of Energy, Office of Science, Office of High Energy 
Physics, under Award DE-SC-0007867 and contract no.\ DE-AC02-05CH11231.


\end{document}